# High-precision Absolute Coordinate Measurement using Frequency Scanned Interferometry

Tianxiang Chen[a, b]    Hai-Jun Yang[c, b,*]    Keith Riles[b]    Cheng Li[a]

[a] State Key Laboratory of Particle Detection and Electronics, University of Science and Technology of China, Hefei 230026, China

[b] Department of Physics, University of Michigan, Ann Arbor 48109, USA

[c] Department of Physics and Astronomy, Shanghai Jiao Tong University, Shanghai 200240, China

[*] Corresponding author. E-mail address: Haijun.Yang@sjtu.edu.cn and yhj@umich.edu (Hai-Jun Yang).

**Abstract:**   In this paper, we report high-precision absolute coordinate measurements performed with frequency scanned interferometry (FSI). We reported previously on measurements of absolute distance with the FSI method [1, 2]. Absolute position is determined by several related absolute distances measured simultaneously. The achieved precision of 2-dimensional measurements is better than 1 micron, and in 3-dimensional measurements, the precision on X and Y is confirmed to be below 1 micron, while the confirmed precision on Z is about 2 microns, where the confirmation is limited by the lower precision of the available translational stage along the 3rd dimension.

## Introduction

Detectors proposed for the International Linear Collider (ILC) [3] will require unprecedented charged-particle track momentum resolutions over a large solid angle. These resolutions require, in turn, precise knowledge of tracking element positions and possible distortions, in order to keep systematic errors due to misalignment to an acceptable level. Here we report frequency scanned interferometry (FSI) measurements carried out as part of an R&D program aimed at aligning the tracking system and embedded accelerator components of the proposed Si detector [4]. These measurements follow previously reported FSI work [1, 2].

In brief, FSI alignment is based on a geodetic grid of point-to-point measurements of distances for lines of sight between beam-splitter reference points and reflector targets. Fitting a large number of absolute distance measurements for an over-constrained grid provides information on the position, orientation and possible distortion of detector elements probed by the FSI beams. Figure 1 depicts schematically one FSI channel, where the distance between the beam splitter (BS) and the retroreflector is determined from counting fringes at the photo-receiver as the frequency of the tunable laser input is scanned over a known range measured by the witness Fabry-Perot interferometer shown. The FSI method was pioneered for high-energy detector alignment by the Oxford ATLAS group [5-7]. More detail on the principles of the FSI method and on the FSI analysis techniques used here can be found in [1], along with a discussion of systematic uncertainties in single-laser, single-channel distance measurements. Discussion of the advantages of a dual-laser FSI method used here and a summary of measurements can be found in [2], but we focus here on consistency checks of a single-laser configuration in a closed, stable environment.

Eventually we wish to apply the FSI method to the alignment of silicon micro-strip layers of the SiD tracking system, its pixel vertex detector and the embedded final-focus quadrupole magnets, using a grid of hundreds of FSI lines of sight. Alignment of the micro-strip layers is critical to precise track momentum resolution; alignment of the vertex detector is critical to precise impact parameter resolution and to displaced vertex reconstruction; alignment of the final-focus magnets is important to establishing collisions of the electron and positron beams in the center of the detector. In order to ensure transverse misalignments of tracking elements to better than 10 μm [8], we aim at longitudinal line-of-sight distance accuracies of 1 μm or better. Previous work demonstrated such precision for single-laser, single-channel distance measurements under highly controlled conditions [1] and for dual-laser, single-channel distance measurements under poorly controlled environmental conditions [2].

Here we present our first single-laser, multiple-channel measurements, in which multiple distance determinations are used to reconstruct 2-dimensional and 3-dimensional coordinates of a stand-in detector element. Under thermostatic and closed box environmental conditions, a precision of about 0.7 microns was achieved in 2-dimensional measurement for a distance of about 24 cm for the prototype. And in 3-dimensional conditions, the confirmed precision of X and Y is about 0.7 microns, while the confirmed precision in Z direction is about 2 microns, with confirmation limited by the low precision of the translational stage in the 3rd dimension.

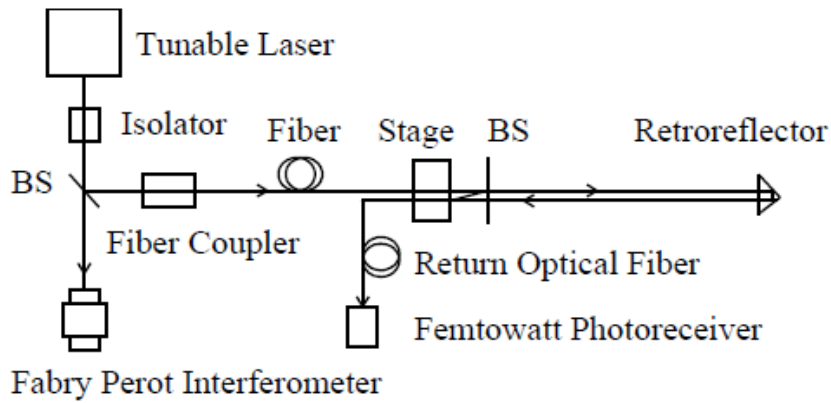

Figure 1: Schematic of a single-channel optical fiber FSI system.

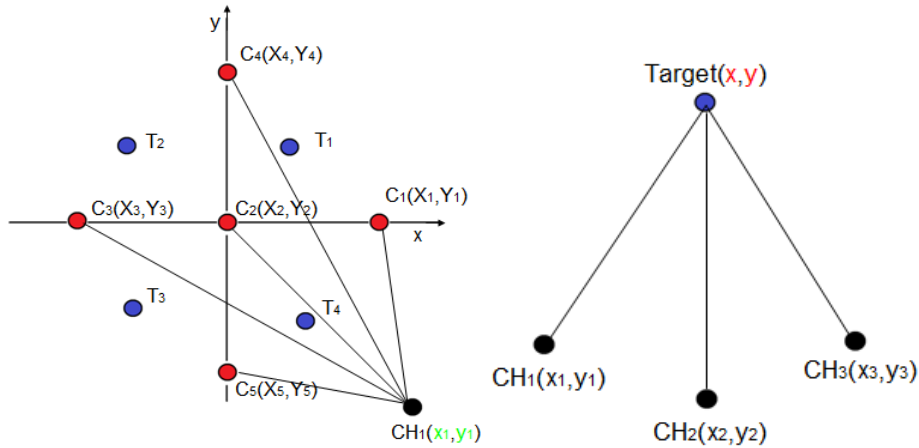

Figure 2: Left, determining the position of fiber-launcher; right, measuring the coordinates of the target. Distances are not shown to scale; the separations between the CHi channels and the targets were much larger than the separations among the target positions.

# Multiple-Channel FSI Demonstration System

For each channel, the absolute distance was defined to be the distance between retroreflector and fiber-launcher. More precisely, we measured the distance between the retroreflector side of the beam splitter and the inside vertex of the hollow corner cube used as the retroreflector. To measure the absolute 2-dimensional coordinates of the retroreflector, we needed first to determine the coordinates of the stationary fiber-launchers. As shown in figure 2, we defined C2 as the origin. We obtained four additional calibrated points by moving the retroreflector to the C1, C3, C4 and C5 positions on a 3-dimensional translation stage. The stage had two orthogonal horizontal motions with digital-readout linear encoders ("Acu-Rite") of precision 0.5 microns defining the x and y directions and a vertical motion with vernier readout of approximately 2-micron precision defining the z direction The absolute distances were taken to be $L_i = \sqrt{(X_i - x_1)^2 + (Y_i - y_1)^2}$, where $X_i$ and $Y_i$ were the measured Acu-Rite translation stage coordinates, $x_1$ and $y_1$ were coordinates of the fiber-launcher CH1 to be determined and i(=1~5) denoted the five calibration points shown in Figure 2. The distances between the fiber-launch CH1 and the calibration points, $FSI_i$, were measured directly using the FSI method described in [1, 2].

We determined the coordinates of the fiber-launcher CH1 by minimizing $\chi^2 = \sqrt{\sum_i^5 (FSI_i - L_i)^2}$ with regard to $(x_1, y_1)$ using a least-squares minimization program (TMinuit function in the CERN ROOT package), where $FSI_i$ was the measured distance between CH1 and Ci. Hence the precision of the calibrated coordinates of CH1 were limited by the precision of the translation stage readout.

After determining the coordinates of the three fiber-launchers (CH1, CH2 and CH3) in the right plot of Figure 2, we could determine the 2-dimensional coordinates of the retroreflector when move to a new location, as long as the displacement did not require readjustment of the fiber launch angle. The demonstration FSI system used is shown in figure 3.

We used a similar method to make 3-dimensional measurements with five fiber launchers. Because of unreliable 2[nd]-laser operation, we report here only single-laser measurements under controlled conditions, using a protective plexiglass enclosure. Otherwise, the measurement equipment and techniques used were similar to those used in [2].

# Two/three-dimensional coordinate measurements with a multiple-channel FSI system

Simultaneous multiple-channel distance measurements are required for practical application of an FSI system to ILC tracker alignment, to determine the 3-dimensional positions of the track elements. To this end, we built the multiple-channel FSI systems for 2-dimensional and 3-dimensional measurements shown in Figure 3. The laser beam was coupled into a single mode optical fiber splitter which split an incoming beam into eight outgoing beams for multiple-channel point-to-point FSI distance measurements. The retroreflector was mounted on the 3-dimensional translation stage, allowing cross check of the displacement of the retroreflector (serving as a stand-in for a detector element).

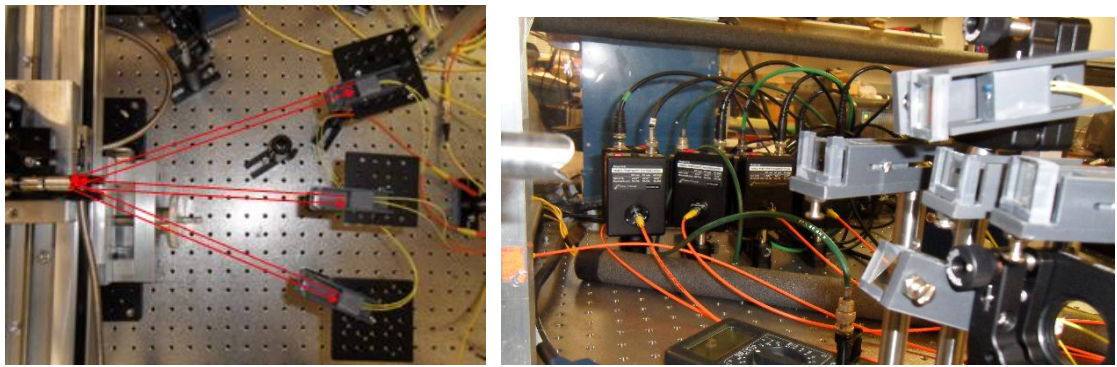

Figure 3: FSI setting for 2-dimensional (left) and 3-dimensional measurements.

Sample results from the 2-dimensional measurements are listed in Table I, where the nominal distance was about 24 centimeters, and the translation stage changed only the 2-dimensional horizontal position of retroreflector. As can be seen, the coordinates reconstructed from the three FSI channels were consistent with the Acu-Rite readouts at each position, for which the displacement measurement precision was about 0.7 microns.

| [microns] | Test1 (x,y) | | Test2 (x,y) | | Test3 (x,y) | | Test4 (x,y) | |
|---|---|---|---|---|---|---|---|---|
| ACU-RITE($\pm\frac{1}{2}\mu$m) | 318 | −291 | −277 | 355 | 318 | 355 | −277 | −500 |
| Reconstruction | 318.7 | −291.6 | −276.5 | 355.8 | 317.3 | 355.7 | −276.1 | −499.8 |
| Difference | −0.7 | 0.6 | −0.5 | −0.8 | 0.7 | −0.7 | −0.9 | −0.2 |

Table I: 2-dimensional coordinate measurements with the 3-channel FSI system

Sample results from the 3-dimensional measurement are listed in Table II, where the nominal distance was about 14 centimeters, and the translation stage changed the 3-dimensional position of the retroreflector. The coordinates of the retroreflector were reconstructed from five FSI channels. As discussed above, the precision of the tuning stage in the X, Y directions was 0.5 microns, while in the Z direction the precision was about 2 microns. As expected, because of the degraded accuracy of translation stage measurements in the vertical direction, there was a noticeably larger spread in vertical displacement residuals with respect to FSI reconstructions.

|  | Test point 1 (x,y,z) | | | Test point 2 (x,y,z) | | | Test point 3 (x,y,z) | | |
|---|---|---|---|---|---|---|---|---|---|
| ACU-RITE(x,y:±½ μm) MICROMETER(z:±2 μm) | −202.5 | 194.5 | 200 | −202.5 | −198 | 200 | 187 | −197.5 | 200 |
| Reconstruction | −203.2 | 195.3 | 201.7 | −202.6 | −197.6 | 201.8 | 186.2 | −197.2 | 202.7 |
| Difference | 0.7 | −0.8 | −1.7 | 0.1 | −0.4 | −1.8 | 0.8 | −0.3 | −2.7 |

Table II: 3-d coordinates measurement with the 3-channel FSI system

## Conclusions

These multiple-channel measurements with 2-dimensional and 3-dimensional FSI demonstration systems confirm that measurements of a set of individual point-to-point distances with the FSI method permit the reliable reconstruction of the fiducial positions of retroreflectors. These measurements were carried out using a single laser in a controlled environment. Based on prior measurements [2] with a dual-laser system in an unfavorable environment, however, we expect these methods and instrumentation to successfully reconstruct detector element displacements in the fluctuating environmental conditions characteristic of detectors in high energy collider interaction regions. Nonetheless, more work will be needed to extend the measurable distance ranges from O(tens cm) to O(meters) needed for alignment of a high energy particle tracking system.

## Acknowledgments

This work was supported by the National Science Foundation and the Department of Energy of the United States.